\documentclass[%
nofootinbib,
 amsmath,amssymb,
pra,
]{revtex4-2}

\usepackage{graphicx}
\usepackage{dcolumn}
\usepackage{bm}
\usepackage{siunitx}  
\usepackage{cleveref}
\usepackage[english]{babel}
\usepackage{lineno}

\let\oldequation\equation
\let\oldendequation\endequation
\let\oldalign\align
\let\oldendalign\endalign

\renewenvironment{equation}
  {\linenomathNonumbers\oldequation}
  {\oldendequation\endlinenomath}

\renewenvironment{align}
  {\linenomathNonumbers\oldalign}
  {\oldendalign\endlinenomath}

\begin{document}

\renewcommand{\Re}[0]{\operatorname{Re}}
\renewcommand{\Pr}[0]{\operatorname{Pr}}
\newcommand{\Sc}[0]{\operatorname{Sc}}
\newcommand{\eps}[0]{\varepsilon}
\newcommand{\nhat}[0]{\widehat{n}}
\newcommand{\br}[1]{\left(#1\right)}

\title{Aspect ratio affects iceberg melting}

\author{Eric W. Hester}
\email{eric.hester@sydney.edu.au}
\affiliation{School of Mathematics and Statistics, University of Sydney, Sydney, NSW 2006, Australia.}

\author{Craig D. McConnochie}
\affiliation{Department of Civil and Natural Resources Engineering, University of Canterbury, Christchurch 8041, New Zealand.}

\author{Claudia Cenedese}
\affiliation{Department of Physical Oceanography, Woods Hole Oceanographic Institution, Woods Hole, MA 02543, USA.}
 
\author{Louis-Alexandre Couston}
\affiliation{British Antarctic Survey, Cambridge CB3 0ET, UK}
\affiliation{Univ Lyon, Ens de Lyon, Univ Claude Bernard, CNRS, Laboratoire de Physique, F-69342 Lyon, France}%

\author{Geoffrey Vasil}
\affiliation{School of Mathematics and Statistics, University of Sydney, Sydney, NSW 2006, Australia.\hspace{1em}}%

\date{\today}

\begin{abstract}
Iceberg meltwater is a critical freshwater flux from the cryosphere to the oceans.
Global climate simulations therefore require simple and accurate parameterisations of iceberg melting.
Iceberg shape is an important but often neglected aspect of iceberg melting.
Icebergs have an enormous range of shapes and sizes, and distinct processes dominate basal and side melting. 
We show how different iceberg aspect ratios and relative ambient water velocities affect melting using a combined experimental and numerical study.
The experimental results show significant variations in melting between different iceberg faces, as well as within each iceberg face. 
These findings are reproduced and explained with novel multiphysics numerical simulations. 
At high relative ambient velocities melting is largest on the side facing the flow, and mixing during vortex generation causes local increases in basal melt rates of over 50\%.
Double-diffusive buoyancy effects become significant when the relative ambient velocity is low.
Existing melting parameterisations do not reproduce our findings.
We propose several corrections to capture the influence of aspect ratio on iceberg melting.
\end{abstract}

\maketitle

\section{Introduction}
Iceberg meltwater provides an important flux of freshwater from ice sheets to oceans \cite{SilvaContributionGiantIcebergs2006,MartinExperimentalTheoreticalStudy1977,HellyCoolingDilutionMixing2011,StephensonSubsurfaceMeltingFreefloating2011},
making up 45\% of Antarctic freshwater loss \cite{RignotIceShelfMeltingAntarctica2013}, 
and dominating freshwater production in Greenland fjords \cite{EnderlinIcebergMeltwaterFluxes2016}.
Melting also releases nutrients that boost biological productivity and carbon sequestration \cite{SmithFreeDriftingIcebergsHot2007}.
Where and when meltwater and nutrients are released depends on how quickly icebergs melt.
Understanding how icebergs influence the climate therefore requires accurate predictions of iceberg melt rates.
We present an experimental and numerical investigation of an often neglected aspect of melting---iceberg shape.

Icebergs display enormous variation in shape and size \cite{BuddAntarcticIcebergMelt1980,TournadreLargeIcebergsCharacteristics2015,VenkateshIcebergLifeExpectancies1988,El-TahanValidationQuantitativeAssessment1987,SilvaContributionGiantIcebergs2006,AndresIcebergsSeaIce2015,RackowSimulationSmallGiant2017}.
Horizontal extents range from several meters to the record iceberg B-15 at $\SI{300}{km} \times \SI{40}{km}$ \cite{ArrigoAnnualChangesSeaice2004}.
Depths vary considerably but almost never exceed \SI{600}{m} \cite{DowdeswellKeelDepthsModern2007}.
Rolling instability further constrains realistic shapes.
Icebergs can tumble when the \emph{aspect ratio}, the ratio of length $L$ to submerged depth $D$, is smaller than $\sqrt{ 
0.92 + {{58.32}}/{D}}$,
where $D$ is expressed in metres \cite{WeeksElementsIcebergTechnology1978,BiggModellingDynamicsThermodynamics1997}.
Aspect ratios may therefore range anywhere from 1 to 1000.
The overall melting will depend strongly on aspect ratio whenever bottom and side melt rates differ.

Using empirical relations for turbulent heat transfer over a flat plate \cite{EckertHeatMassTransfer1959},
Weeks and Campbell developed a commonly used parameterisation for iceberg melt rates \cite{WeeksIcebergsFreshWaterSource1973} (hereafter the WC model).
The WC model predicts an iceberg melt rate $v$ (in dimensional units of speed) of
	\begin{equation}
		v_{WC} = 0.037 \left(\frac{\rho_w}{\rho_i} \nu^{-7/15}\kappa^{2/3} \frac{c_p}{\Lambda}\right) \frac{U^{0.8}\Delta T}{L^{0.2}}.
		\label{eq:Weeks_Campbell}
	\end{equation}
Here, $\rho_w$ and $\rho_i$ are the respective densities of water and ice, $\nu$ and $\kappa$ are the respective diffusivities of momentum and temperature, $c_p$ is the heat capacity at constant pressure of seawater, $\Lambda$ is the latent heat of ice melting, $U$ is the relative speed between the ambient water and the iceberg, $L$ is the iceberg length, and $\Delta T$ is the difference between ambient and melting temperatures.

Unfortunately the WC model makes several incorrect predictions.
One erroneous prediction--zero melting at zero relative velocity--was recently addressed by FitzMaurice et al. \cite{FitzMauriceNonlinearResponseIceberg2017}, by considering meltwater-plume entrainment for low relative ambient velocities.
Their parameterisation (hereafter the FM model) is equivalent to the WC model when ambient speeds $U$ are greater than the plume speed $U_p$.
For slower ambient flows, the FM model represents entrainment by replacing the flow speed $U$ with the plume speed $U_p$, and multiplying the water temperature $T_w$ by the factor $((1 + U^2/U_p^2)/2)^{1/2}$.
Yet the WC and FM models still miss an important point--side and base melting can differ.
Buoyancy forces, which depend on water temperature and salinity, generate different flows at the base and side of the iceberg.\footnote{
We further note that iceberg sides also deteriorate via wave erosion and calving  \cite{BiggModellingDynamicsThermodynamics1997}, though this investigation focusses on melting.}

A second common parameterisation of ice shelf melting was developed by Holland and Jenkins \cite{HollandModelingThermodynamicIce1999}.
Their ``three equation'' model (hereafter the HJ model) solves for the melting temperature and heat and salt conservation to determine the interface temperature $T_b$, salinity $C_b$ and melt rate $v$,
	\begin{subequations}
	\begin{align}
		T_b &= \lambda_1 C_b + \lambda_2,\\
		v\, (\rho_i \Lambda + \rho_i c_i (T_b - T_i)) &=  \rho_w c_p C_d^{1/2} U \Gamma^T (T_w - T_b),\\
		v\, \rho_i C_b &= \rho_w C_d^{1/2} U \Gamma^C ( C_w - C_b ).
	\end{align}
	\end{subequations}
In general the interface temperature and salinity differ from their far-field values in the ice ($T_i,C_i$) and water ($T_w,C_w$).
The coefficients $\lambda_1$ and $\lambda_2$ represent a linearised approximation to the melting temperature, and heat and salt fluxes are parameterised in terms of the flow speed $U$ through transfer coefficients $\Gamma^{T}$ and $\Gamma^{C}$ and a drag coefficient $C_d$.
While commonly used, this parameterisation is completely independent of iceberg geometry. 

Our goal is to understand the effect of aspect ratio on iceberg melting in a series of laboratory experiments and numerical simulations.
We compare our findings with predictions of the WC, FM, and HJ models and suggest improvements to account for the influence of aspect ratio on melting.
In section \ref{sec:experimental-methods} we describe the experimental method, and summarise the findings in section \ref{sec:experiment-results}.
Using a recent numerical method \cite{HesterIcebergmeltingcode2020}, summarised in section \ref{sec:comp-methods}, we reproduce the laboratory experiments in a series of fluid-solid simulations, which allow us to identify key physical processes controlling melting in section \ref{sec:comp-results}. 
We summarise geophysical implications and discuss possible improvements to parameterisations in section \ref{sec:geophysical}.
We conclude and discuss future directions in section \ref{sec:conclusion}.

\section{Experimental Methods}
\label{sec:experimental-methods}
	\begin{figure}[ht]
	\begin{center}
	\includegraphics[width=.5\linewidth]{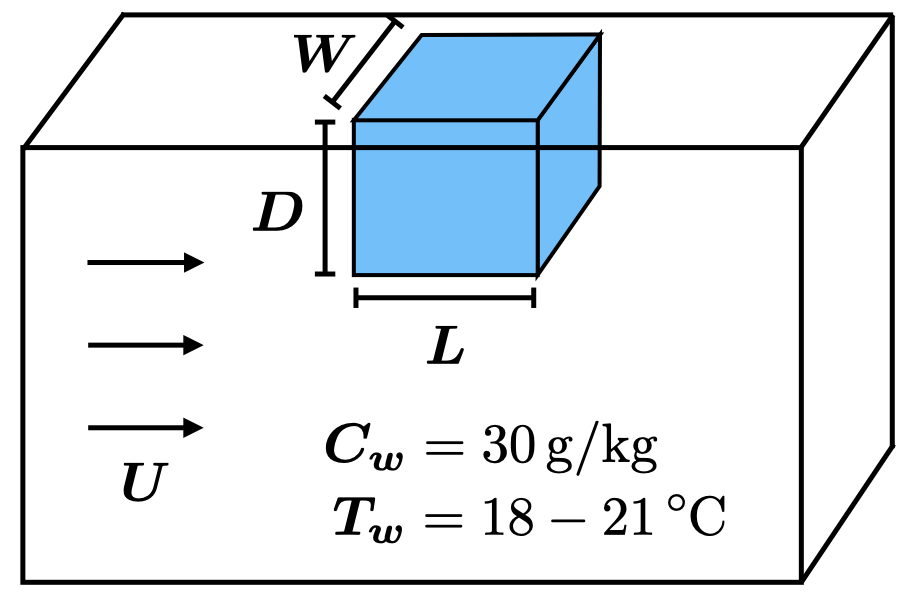}
	    \caption[Initial photo]{
    Experimental schematic. A dyed ice block of dimensions with length $L$, width $W$, and immersed depth $D$ is fixed in a flume with a relative ambient velocity $U$, temperature $T_w$ and salinity $C_w$. }
    \label{fig:tank}
	\end{center}
	\end{figure}

The experiments immersed ice blocks of different lengths in a recirculating salt water flume with different ambient water velocities (\cref{fig:tank}).
The central section of the flume measured \SI{76.5}{cm} long by \SI{42}{cm} wide by \SI{33.5}{cm} deep, and recirculated salt water with salinity $C_w = \SI{30}{g/kg}$ and temperatures $T_w = 18$ to $\SI{21}{\celsius}$.
The 26 distinct experiments held the depth fixed at $D = \SI{3}{cm}$. 
We considered 5 lengths $L = 10, 15, 20, 25, 33$ cm and three relative ambient flow velocities $U = 0, 1.5$, or \SI{3.5}{cm.s^{-1}}.
The width also varied from $W = 10$ to \SI{22.5}{cm}, but did not affect melt rates.
We therefore do not comment further on it.
The water of each ice block was dyed blue and left to de-gas overnight before freezing at $T_i = -\SI{30}{\celsius}$.
The ice block was weighed, positioned in the flume, and each experiment ran for 10 minutes.
The ice block was then removed, reweighed to calculate mass loss, and photographed from each side.

Post-experiment photographs determined the melting of each face.
A blue filter highlighted the block, then thresholding returned a binary image, and an opening transform removed the noise.
The edges of the waterline are defined as the lowest points with less than 15 pixels of melting, and the images were rotated to level the waterline.
Pixels were converted to cm using rulers in each photo, and the left (front) profile was calculated as a function of depth.
To distinguish the faces, the left face was defined as the portion of the outline with slope greater than 1.
The depth average of the left face determined the average melt rate of that face.
Average bottom and right melt rates were calculated similarly.

\section{Experimental results}
\label{sec:experiment-results}

\subsection{Qualitative observations}
\textbf{High relative ambient velocity} ---
	\begin{figure}[t]
	\centering
    \begin{center}
    \includegraphics[width=.9\linewidth]{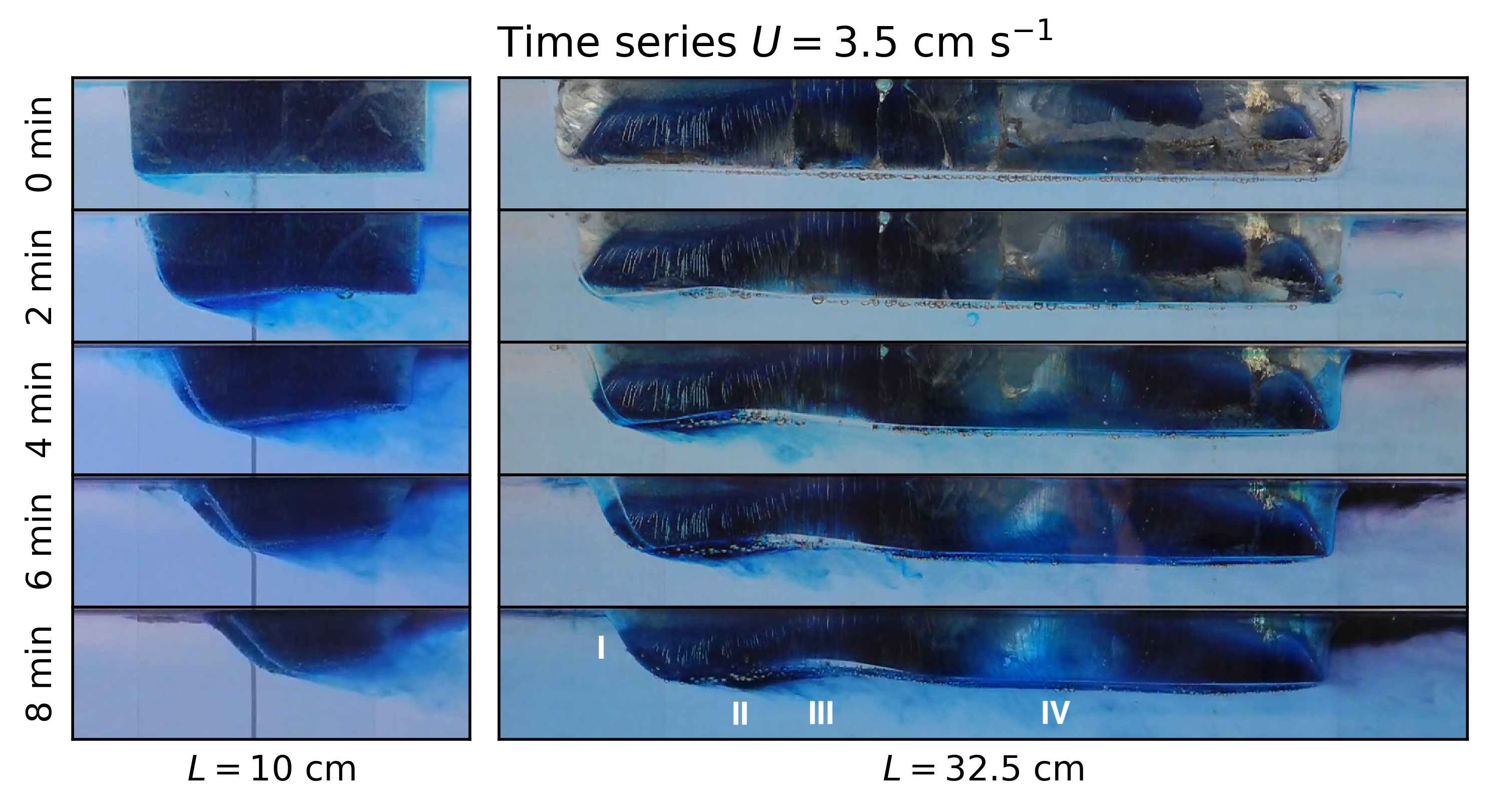}
    \caption[Time series]{Time series of two experiments with $L = \SI{10}{cm}$ (left) and $L = \SI{32.5}{cm}$ (right).
    Each block was immersed to $D = \SI{3}{cm}$, in ambient water moving at $U=\SI{3.5}{cm\,s^{-1}}$.
    Each frame is separated by two minutes.}
    \label{fig:fast-time-series} 
    \end{center}
	\end{figure}
\Cref{fig:fast-time-series} shows a time series of two experiments at $U = \SI{3.5}{cm\,s^{-1}}$.
The melting on both front faces (left side of each frame) is much larger than on the base and sides.
This result agrees with previous studies showing that flow perpendicular to an ice face increases heat transport and melt rates compared to flow parallel to an ice face \cite{JosbergerLaboratoryTheoreticalStudy1981}.
We thus expect aspect ratio to influence melting at large relative ambient flow speeds.

The second apparent feature is the non-uniformity of melting within each face of each experiment (\cref{fig:fast-time-series}).
The front melting increases with depth (\cref{fig:fast-time-series} I), decreasing the slope of the leading face over time.
The basal melting also has a pronounced non-uniform profile.
Starting from upstream, a darker region of increased dye concentration is pooled just behind the front of the base, and the melt rate is low (\cref{fig:fast-time-series} II).
This concentration suggests a stagnant zone that does not mix with the incoming flow, which is typical when flow separation occurs.
Immediately behind this region we observe increased turbulence and basal melt rates in both experiments (\cref{fig:fast-time-series} III).
At the rear the longer ice block (\cref{fig:fast-time-series}, right) returns to a regime of lower, uniform melting (\cref{fig:fast-time-series} IV).
The dye patterns toward the rear of the longer block (right) become less mixed, suggesting less turbulent flow.
The melting pattern is similar for all blocks, and echoes understanding that other liquid-solid phase change problems can evolve to self-similar shapes \cite{HaoHeatTransferCharacteristics2002,MooreSelfsimilarEvolutionBody2013,HuangShapeDynamicsScaling2015}.
The lateral sides of the block (facing the reader in \cref{fig:fast-time-series}) exhibited similar local increases in melting in this region, though they were less pronounced.

The flow field helps explain the basal melting pattern.
As fluid moves past a forward-facing step, vorticity separates from the leading edge.
This configuration produces a region of unsteady recirculation and subsequent reattachment of the flow.
Studies examining heat transfer in flow past a forward-facing step find a maximum in convective heat transfer at the point of reattachment \cite{Abu-MulawehTurbulentMixedConvection2005,JayakumarHybridMeshFinite2010}.
Different estimates for the reattachment length exist (summarised in \cite{SherryFlowSeparationCharacterisation2009}) but all find that it is roughly 3 to 5 times the step height for Reynolds numbers $\Re = 10^3$ to $10^5$ (our experiment is at $\Re \equiv UD/\nu = 800$, based on the step height).
Actual icebergs in Greenland fjords may experience Reynolds numbers up to \num{2e7}, assuming a draft of \SI{200}{m}, and local relative velocities up to \SI{0.1}{m.s^{-1}} \cite{FitzMauriceEffectShearedFlow2016}.
We predict the reattachment length remains proportional to depth at large scales,
and therefore expect similar localised increases in basal melting for real icebergs.

The free surface behind the large block is also much darker than for the short block. 
While more meltwater is expected for the longer block due to its greater surface area, the fluid flow patterns also help explain this observation.
The flow underneath the shorter block is entirely turbulent, thoroughly mixing the meltwater.
The flow underneath the longer block is instead more laminar, indicating that most meltwater does not mix with the ambient water and rises to the free surface.
The increased turbulence at the rear of the short block also explains the greater melt rate of its rear face.

	\begin{figure}
    \begin{center}
    \includegraphics[width=\linewidth]{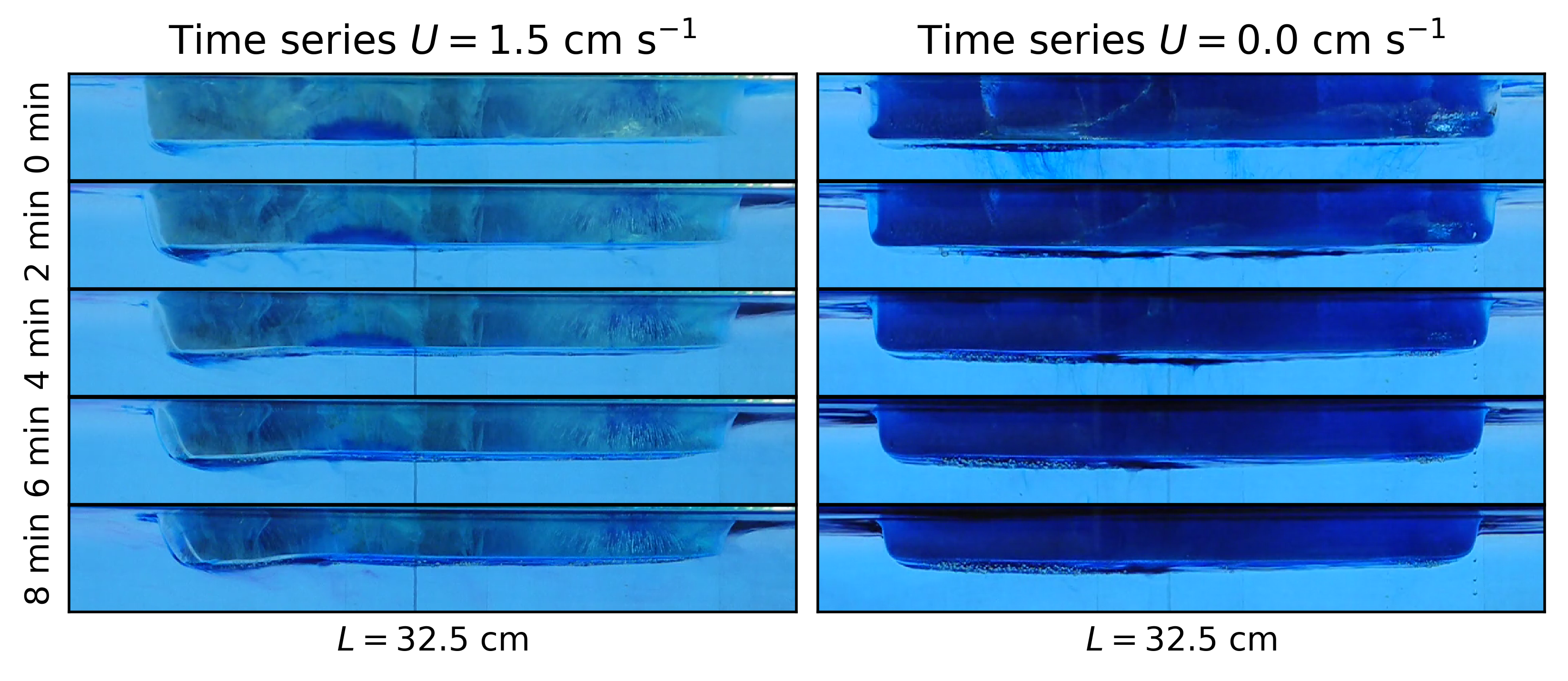}
    \caption[Time series]{Time series of two $L = \SI{32.5}{cm}$ experiments with $U= \SI{1.5}{cm.s^{-1}}$ (left) and $U = \SI{0}{cm\,s^{-1}}$ (right).}
    \label{fig:slow-time-series}
    \end{center}
	\end{figure}

\textbf{Low relative ambient velocity} ---
The left column of \cref{fig:slow-time-series} shows a time series from an experiment performed at \SI{1.5}{cm\,s^{-1}}.
Many of the same trends are apparent from the $U=\SI{3.5}{cm\,s^{-1}}$ experiments.
The basal melt is at a maximum behind the leading (left) edge, followed by a return to laminar flow and more uniform melt rate further downstream.
The stagnant region size is comparable to that in the higher relative velocity experiments, which supports a $\Re$ independent scaling of reattachment length \cite{SherryFlowSeparationCharacterisation2009}.
The primary difference between the $U=1.5$ and \SI{3.5}{cm\,s^{-1}} experiments is lower overall melting and reduced turbulence (dye streaks appear mostly laminar).
Meltwater preferentially pools near the free surface, as there is less mixing with ambient water than at higher relative velocities $U$ (\cref{fig:fast-time-series}).

\textbf{No relative ambient velocity} ---
The zero relative velocity experiments lack any local increases in the melt rate.
Most meltwater flows slowly along the base of the ice block before rising to the surface.
However some dyed fluid sinks from the base of the ice as a dense plume.
This is best understood as a double-diffusive effect, resulting from the lower diffusivity of salinity in comparison to temperature \cite{GadeMeltingIceSea1979}.
During the melting, the adjacent salt water absorbs latent heat.
The cooling occurs faster than salinity diffusion.
With weak ambient flow, the saltwater can cool enough to sink and entrain dyed meltwater along with it.

\subsection{Quantitative results}

	\begin{figure}[t]
    \begin{center}
    \includegraphics[width=\linewidth]{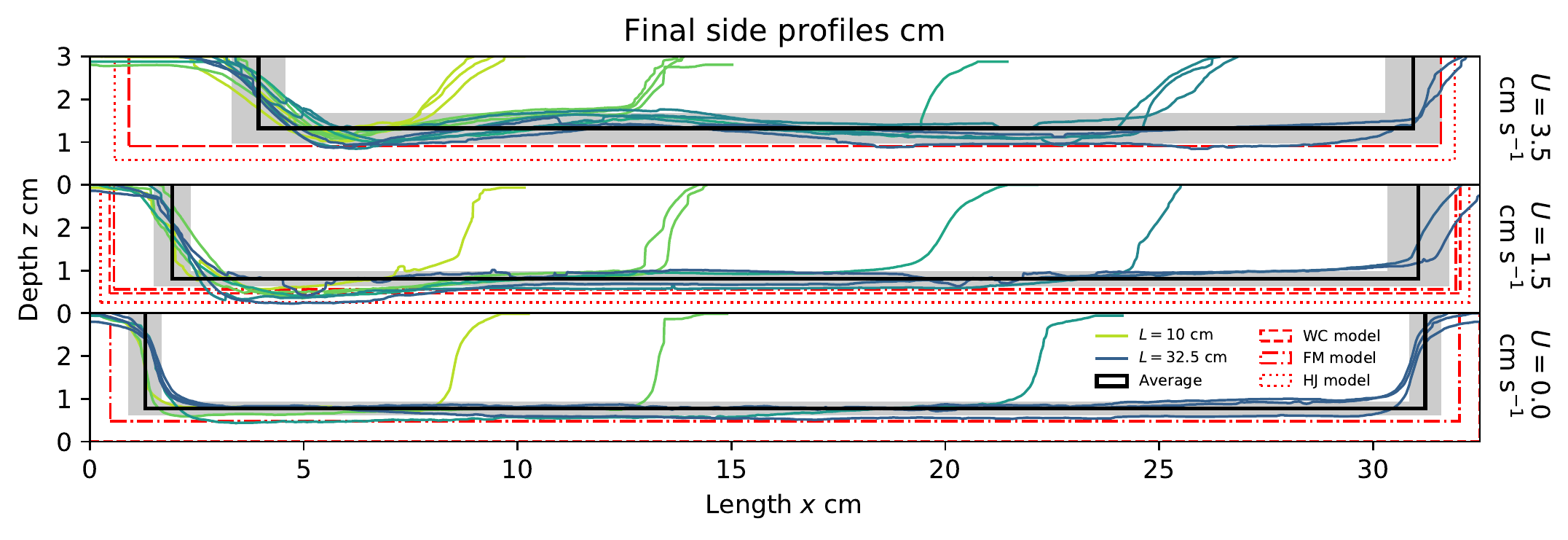}
    \caption{The top, middle, and bottom figures plot final side profiles of each experiment relative to the bottom of the front face at the start of each experiment (origin), for ambient velocities $U={3.5},1.5,\SI{0}{cm.s^{-1}}$.
    Profiles are coloured according to initial aspect ratio.
    The solid black rectangle represents the average total melt of an initially $\SI{32.5}{cm}\times\SI{3}{cm}$ block.
    The average melt is estimated from averages of each face averaged over all experiments for each fluid velocity $U$.
    The grey bars represent twice the standard deviation in average melt rate between experiments.
    The dashed, dashdotted, and dotted red rectangles plot melt rates predicted by the WC, FM, and HJ models respectively.
    All melt rates and uncertainties are given in \cref{tab:melts}.
     }
    \label{fig:profiles}
    \end{center}
	\end{figure}

\Cref{fig:profiles} plots the post-experiment profile of each ice block, and supports previous observations.
Melt rates differ between faces, with lowest melting on the base, and highest melting on the sides.
Larger ambient velocities $U$ cause larger melt rates, particularly for the front face.
Significant non-uniformities in melt rate exist on each face.
These non-uniformities are similar for different experiments at the same ambient velocity $U$.

\Cref{fig:profiles} also illustrates the average melt rates in black, with uncertainties shown in grey.
The black rectangle represents the final shape of a hypothetical block with initial dimensions \SI{32.5}{cm} long and \SI{3}{cm} deep.
The melt rates of each face are calculated by first averaging the melt over each face, then averaging these face melt rates over all experiments at a given ambient velocity $U$.
The uncertainties (grey bars) are defined as twice the standard deviation of the average melt rates between experiments.
The dashed, dashdotted, and dotted red rectangles plot melt rates predicted by the WC, FM, and HJ models respectively.
All models significantly underestimate melting, and ignore non-uniformities within and between faces, though the FM model does predict melting without ambient flows.
	\begin{table}[h!]
	\begin{ruledtabular}
	\begin{center}
    \begin{tabular}{c|ccccccc}
    	Velocity	& Front melt rate & Side melt rate & Rear melt rate& Basal melt rate & WC melt rate & FM melt rate & HJ melt rate \\
    	$U$ \si{cm.s^{-1}} & $v_f$ \si{cm.min^{-1}} & $v_s$ \si{cm.min^{-1}} & $v_r$ \si{cm.min^{-1}} & $v_b$ \si{cm.min^{-1}} & $v_{WC}$ \si{cm.min^{-1}}  & $v_{FM}$ \si{cm.min^{-1}} & $v_{HJ}$ \si{cm.min^{-1}}\\[.1em]\hline 
    	${3.5}$ & $0.39 \pm 0.06$ & $0.21 \pm 0.05$	& $0.16 \pm 0.07$ & $0.13 \pm 0.04$  & $0.091-0.115$ & $0.091-0.115$ & $0.058$ \\
    	${1.5}$ & $0.19 \pm 0.04$ & $0.15 \pm 0.04$	& $0.14 \pm 0.07$ & $0.08 \pm 0.018$ & $0.047-0.058$ & $0.056-0.071$ & $0.025$\\
    	${0.0}$	& $0.13 \pm 0.04$ & $0.15 \pm 0.04$	& $0.13 \pm 0.04$ & $0.08 \pm 0.016$ & 0 & $0.048-0.060$ & 0 
    \end{tabular}
    \caption{Average melt rates of each face for each flow speed $U$ from \cref{fig:profiles}, and WC, FM, and HJ model predictions. 
    The experimental uncertainties are twice the standard deviation in average melt rates in \cref{fig:profiles}.
    The WC model melt rates are given for upper and lower lengths $L_{max}=\SI{32.5}{cm}$ and $L_{min} = \SI{10}{cm}$,
	with
	$\rho_i=\SI{0.9167}{g.cm^{-3}}$,
	$\rho_w=\SI{1.021}{g.cm^{-3}}$,
	${\nu=\SI{1.00e-2}{cm^2.s^{-1}}}$,
	$\kappa=\SI{1.42e-3}{cm^2.s^{-1}}$,
	$c_p = \SI{4.182}{J.g^{-1}.\celsius^{-1}}$,
	$\Lambda = \SI{334}{J.g^{-1}}$, and
	$\Delta T = \SI{20}{\celsius}$, 
	appropriate for ambient water at \SI{20}{\celsius} \cite{BatchelorIntroductionFluidDynamics2000}.
	The FM model uses the same parameters with a plume velocity $U_p$ of \SI{2.4}{cm.s^{-1}} \cite{FitzMauriceNonlinearResponseIceberg2017}.
	The HJ model also uses
	$T_i = -\SI{15}{\celsius}$,
	$c_{p,i} = \SI{2.108}{J.g^{-1}.\celsius^{-1}}$, 
	$C_d = 0.0097$, 
	$\Gamma_T = 0.011$, 
	$\Gamma_C = 3.1e-4$, 
	$\lambda_1 = \SI{-0.057}{\celsius.kg/g}$, 
	$\lambda_2 = \SI{0.083}{\celsius}$ from \cite{JenkinsObservationParameterizationAblation2010}.}
    \label{tab:melts}
    \end{center}
	\end{ruledtabular}
	\end{table}

These numerical values of the average melt rates and uncertainties are reproduced in \cref{tab:melts}.
The second through fifth columns of \cref{tab:melts} give experimental melt rates for the front, side, rear and basal melt rates respectively, at each relative velocity $U$.
The melt rates for each face differ at all relative velocities $U$.
Aspect ratio can therefore affect overall melting by changing the relative areas of these faces.
The melt rate of each face increases slightly from $U = \SI{0}{cm.s^{-1}}$ to $U=\SI{1.5}{cm.s^{-1}}$, and significantly from $U=\SI{1.5}{cm.s^{-1}}$ to the more turbulent \SI{3.5}{cm\,s^{-1}} experiments.
This result agrees with \cite{FitzMauriceNonlinearResponseIceberg2017}, which observed roughly constant melt rates below a threshold relative fluid velocity of \SI{2.4}{cm.s^{-1}} in similar laboratory experiments with aspect ratio less than one.
While FitzMaurice et al.~focussed on side melting via plume entrainment, and did not measure independent melting for each face, the existence of a similar threshold velocity for basal melting is likely.

We give estimated melt rates using the WC, FM, and HJ models\footnote{
	Note that we do not use the average internal ice temperature of $T_i = \SI{-15}{\celsius}$ (the midpoint of the melting temperature \SI{0}{\celsius} and freezer temperature \SI{-30}{\celsius}) for the WC and FM models, as was done in \cite{FitzMauriceNonlinearResponseIceberg2017}. 
	Using an internal ice temperature implies that colder blocks will melt faster.
	This contradicts the fact that colder blocks reduce the heat flux difference at the interface, and therefore melt more slowly.
	We instead use the physically justified melting temperature of the interface, giving $\Delta T = \SI{20}{\celsius}$.} 
	in the rightmost column of \cref{tab:melts}.
The upper and lower values for the WC and FM models are for the longest ($L=\SI{32.5}{cm}$) and shortest ($L=\SI{10}{cm}$) block lengths.
All models underestimate melt rates of all faces at all velocities, and the disagreement worsens for lower velocities, as found in \cite{FitzMauriceNonlinearResponseIceberg2017}.
The only estimate within experimental uncertainties is the WC and FM model prediction for basal melt rates at $U=\SI{3.5}{cm.s^{-1}}$.
Importantly, no model predicts different melt rates for different faces.
An accurate parameterisation must therefore account for both magnitude and orientation of relative ambient flow.

\section{Computational methods}
\label{sec:comp-methods}
Existing melting parameterisations do not agree with our experimental findings.
We use Direct Numerical Simulation (DNS) of melting ice in warm salty water to investigate the full flow dynamics and further our understanding of our laboratory observations.
We simulate this problem with a recent phase-field approach developed for coupled fluid flow, melting, and dissolution \cite{HesterImprovedPhasefieldModels2020}.
This method builds on previous methods for fluid-solid interactions \cite{HesterImprovingAccuracyVolume2020} and simulations of melting in fresh water \cite{CoustonTopographyGenerationMelting2020,PurseedBistabilityRayleighBenardConvection2020}.

\subsection{Phase-field model of ice in warm salt water}
\label{sec:phase-field}
Ice melting in salt water is often modelled as a moving boundary problem \cite{WorsterSolidificationFluids2002}.
In this formulation, the problem is first subdivided into fluid and solid domains.
In the fluid, the temperature $T$ and salinity $C$ satisfy advection-diffusion equations, 
and the fluid velocity ${u}$ and pressure $p$ satisfy incompressible Navier-Stokes equations with vertical buoyancy forcing $-g \rho(T,C) \hat{z}$, where $g$ is acceleration due to gravity.
The solid temperature follows a diffusive equation.
The moving boundary formulation is completed with boundary conditions at the melting interface.
The temperature is continuous and equal to the melting temperature, and Stefan, Robin, and Dirichlet boundary conditions conserve energy, salt, and mass respectively.

Phase-field models are a smoothed approximation of the moving boundary formulation that is physically motivated and simple to simulate \cite{BeckermannModelingMeltConvection1999}.
Distinct phases are represented with a smooth \emph{phase field} $\phi$ that is forced to $\phi\approx 1$ in the solid and $\phi\approx 0$ in the fluid. 
A thin region of size $\eps$ separates the fluid and solid.
The phase-field equations augment the bulk equations with smooth source terms that reproduce the boundary conditions in the limit $\eps\to 0$,
\begin{subequations}
    \begin{align}
	\label{eq:phase-pdes}
    \eps\frac{5}{6}\frac{\Lambda}{c_p\kappa}\partial_t\phi -\gamma \nabla^2 \phi + \frac{1}{\eps^2}\phi(1-\phi)(\gamma (1 - 2\phi) +\eps(T + \lambda C)) &= 0,\\
    \partial_t T + \nabla \cdot \left((1-\phi) u T - \kappa \nabla T\right) &= \frac{\Lambda}{c_p} \partial_t \phi ,\\
    \partial_t((1-\phi+\delta) C) + \nabla \cdot \left( (1-\phi+\delta)(u C - \mu \nabla C\right)) &= 0,\\
    \partial_t{u}+ {u} \cdot\nabla{u}- \nu \nabla^2{u}+ \nabla p + \frac{g\rho(T,C)}{\rho_0}{\hat{z}} &= - \frac{\nu}{\eta} \phi \,{u},\\
    \nabla \cdot{u}&= 0.
    \end{align}
\end{subequations}
Here $\nu, \kappa,$ and $\mu$ are momentum, thermal, and salt diffusivity, $\Lambda$ is the latent heat, $c_p$ is the heat capacity of water, and $\lambda$ is a slope coefficient of the liquidus, which is assumed linear for simplicity.
The damping time $\eta\ll 1$ suppresses advection in the solid \cite{HesterImprovingAccuracyVolume2020}, $\gamma \ll 1$ expresses curvature dependence of the melting temperature, and $\delta \ll 1$ regularises the salinity equation within the ice \cite{HesterImprovedPhasefieldModels2020}.
Replacing moving boundary conditions with smooth source terms allows simple numerical implementations that converge to the moving boundary formulation with error of order $\eps^2$ \cite{HesterImprovedPhasefieldModels2020}.

Note that this model omits second order thermodynamic effects.
We ignore density changes during melting, consider constant viscous, thermal, and solutal diffusivities, and describe buoyancy with the Boussinesq approximation (using the EOS-80 equation of state of seawater \cite{FofonoffAlgorithmsComputationFundamental1983}).
We also use two-dimensional simulations to reduce computational costs.
While the experiments did not vary much in the spanwise direction orthogonal to the flow, the inverse energy cascade in two dimensions is known to generate larger vortices than in three dimensions.

\subsection{Numerical method and simulation parameters}
We perform two series of simulations.
The first series considers large relative ambient velocity $U = \SI{3.5}{cm.s^{-1}}$, and compares a simulation with temperature, salinity,
and buoyancy effects turned on ($T+C$) and a simulation with salinity and buoyancy forcing neglected ($T$ only).
The second series investigates double-diffusive effects with no relative ambient velocity $U = \SI{0}{cm.s^{-1}}$ by comparing simulations with equal temperature and salt diffusivities (single diffusion: SD) and different temperature and salt diffusivities (double diffusion: DD).
All simulations specify the initial ice temperature as $T_i = \SI{0}{\celsius}$, the initial water temperature as $T_w = \SI{20}{\celsius}$, the initial salinity as $C_w = \SI{30}{g.kg^{-1}}$, the liquidus slope as $\lambda = \SI{0.056}{\celsius\,kg\,g^{-1}}$, the latent heat as $\Lambda = \SI{3.34e2}{J\,g^{-1}}$, and the heat capacity as $c_p = \SI{4.2}{J.g^{-1}.\celsius^{-1}}$ \cite{JenkinsObservationParameterizationAblation2010}.
The relative ambient velocity $U$, domain dimensions $\ell\times d$, block dimensions $L\times D$, and diffusivities $\nu,\kappa,\mu$ are given in \cref{tab:simulation-params-model}.
We use a realistic Prandtl number $\operatorname{Pr} = \nu/\kappa$ of 7.
Computational constraints limit feasible salt diffusivities, so the Schmidt number $\Sc = \nu/\mu$ is at most 50 rather than 500.

All simulations are performed using the spectral code Dedalus \cite{BurnsDedalusFlexibleFramework2020}.
The equations are discretised with Fourier series in the horizontal direction, and trigonometric series in the vertical direction.
This corresponds to periodic horizontal boundary conditions; homogeneous Neumann vertical boundary conditions for the horizontal velocity $u$, temperature $T$, salinity $C$, pressure $p$, and phase field $\phi$; and homogeneous Dirichlet vertical boundaries for the vertical velocity $w$.
The simulations at $U=\SI{3.5}{cm.s^{-1}}$ use a \SI{2}{cm} thick volume penalised `sponge layer' at the left of the domain to force the fluid temperature, salinity, and velocity to ambient values \cite{HesterImprovingAccuracyVolume2020}.
The initially stationary liquid is thereby accelerated to the ambient value over the course of a second.
The parity constraint of the buoyancy term in the vertical momentum equation is enforced by tapering buoyancy near the vertical boundaries over a length scale of $\SI{0.05}{cm}$.
Simulations using more expensive Chebyshev discretisations did not taper the buoyancy, but gave essentially identical results.
The system is integrated using a second order implicit-explicit Runge-Kutta timestepper.
The horizontal mode number $n_x$, vertical mode number $n_z$, time step size $\Delta t$ and numerical phase field parameters are given in \cref{tab:simulation-params-numerical}.
The time step is chosen according to the stiff time scale induced by the phase field equation, which is more stringent than the CFL condition.
Owing to computational constraints, the flow simulations are run until $t = \SI{180}{s}$, while the cheaper no-flow simulations are run until $t = \SI{360}{s}$.
All code is available online \cite{HesterIcebergmeltingcode2020}.

	\begin{table}[hbt]
	\begin{ruledtabular}
	\centering
	\caption{Model parameters for flow and no-flow simulations.}
	\label{tab:simulation-params-model}
	\begin{tabular}{cccccccccc}
    Simulation & $U$ & $\ell$ & $d$ & $L$ & $D$ & $\nu$ & $\kappa$ & $\mu$ & $\Sc$\\
    	& \SI{}{cm.s^{-1}} & \SI{}{cm} & \SI{}{cm} & \SI{}{cm} & \SI{}{cm} & \SI{}{cm^2.s^{-1}} & \SI{}{cm^2.s^{-1}} & \SI{}{cm^2.s^{-1}} & \\
    \hline
    Flow $T+C$ & 3.5 & 50 & 15 & 30 & 3 & \num{1.3e-2} & \num{1.86e-3} & \num{9.3e-4} & 14\\
    Flow $T$ only & 3.5 & 50 & 15 & 30 & 3 & \num{1.3e-2} & \num{1.86e-3} & N/A & N/A\\
    No-flow DD & 0 & 20 & 10 & 10 & 3 & \num{1.3e-2} & \num{1.86e-3} & \num{2.6e-4} & 50\\
    No-flow SD & 0 & 20 & 10 & 10 & 3 & \num{1.3e-2} & \num{1.86e-3} & \num{1.86e-3} & 7 
	\end{tabular}
	\end{ruledtabular}
	\end{table}

	\begin{table}[hbt]
	\begin{ruledtabular}
	\centering
	\caption{Numerical parameters for flow and no-flow simulation series.}
	\label{tab:simulation-params-numerical}
	\begin{tabular}{ccccccccc}
    Simulation & $n_x$ & $n_z$ & $\Delta t$ & $\eps$ & $\gamma$ & $\eta$ & $\delta$\\
    series & & & \SI{}{s} & \SI{}{cm} & \SI{}{cm.\celsius} & \SI{}{s} & \\
    \hline
    Flow    & 6144 & 1536 & \num{1.6e-4} & 0.01 & 0.2 & \num{1.75e-3} & \num{5e-3}\\
    No-flow & 2048 & 1024 & \num{5e-4}   & 0.01 & 0.2 & \num{1.75e-3} & \num{1e-4}
	\end{tabular}
	\end{ruledtabular}
	\end{table}
	
\section{Computational results}
\label{sec:comp-results}
\subsection{High relative ambient velocity}

	\begin{figure}
	\centering
	\includegraphics[width=\linewidth]{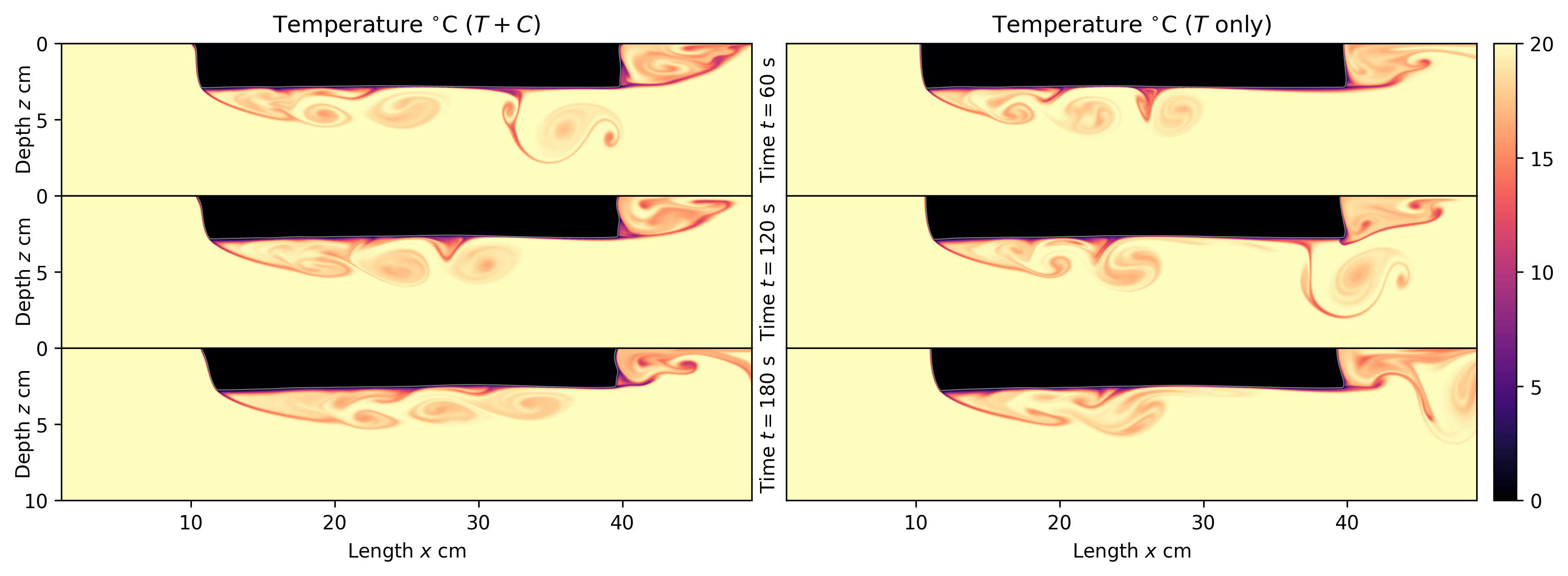}
	\caption{Time series, at one minute intervals, of temperature for melting simulations in warm salt water at relative ambient velocity $U = \SI{3.5}{cm.s^{-1}}$.
	The $T+C$ simulation (left) includes temperature, salinity, and buoyancy, while the $T$ only simulation (right) omits salinity and buoyancy effects.}
	\label{fig:time-series-flow}
	\end{figure}
	
	\begin{figure}
	\includegraphics[width=.9\linewidth]{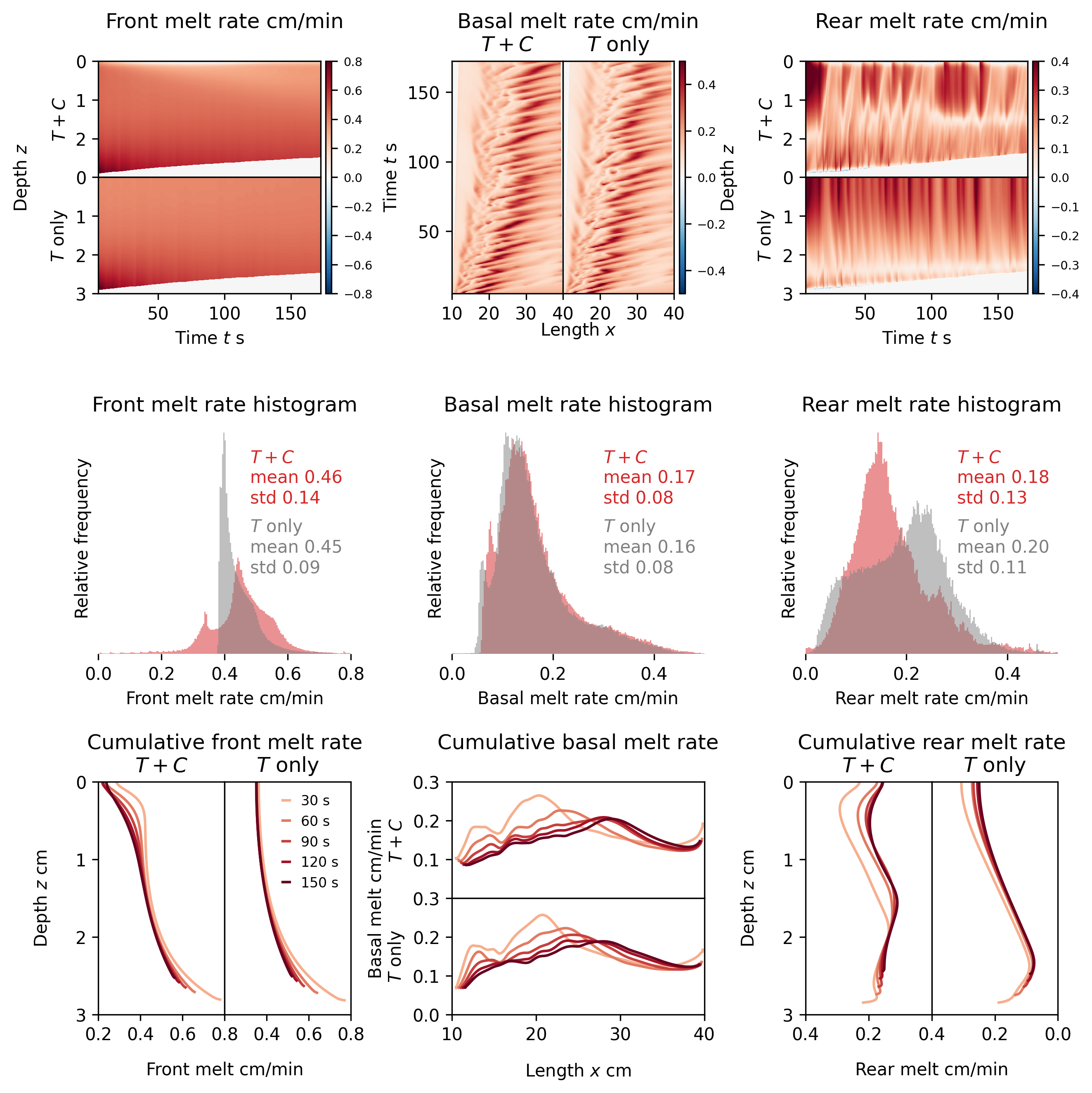}
	\caption{The left, middle, and right columns summarise front, base, and rear melting respectively for $U=\SI{3.5}{cm.s^{-1}}$ simulations ($T+C$ and $T$ only).
	The front is defined where the interface slope is less than $-1$, the base where the slope is between $-1$ and 1, and the rear where the slope is greater than 1.
	The top row plots instantaneous melt rates in space and time.
	The centre row gives histograms, mean, and standard deviation of instantaneous melt rates.
	The bottom row plots cumulative melt rates over time.}
	\label{fig:melting-flow}
	\end{figure}

	\begin{figure}[h]
	\includegraphics[width=\linewidth]{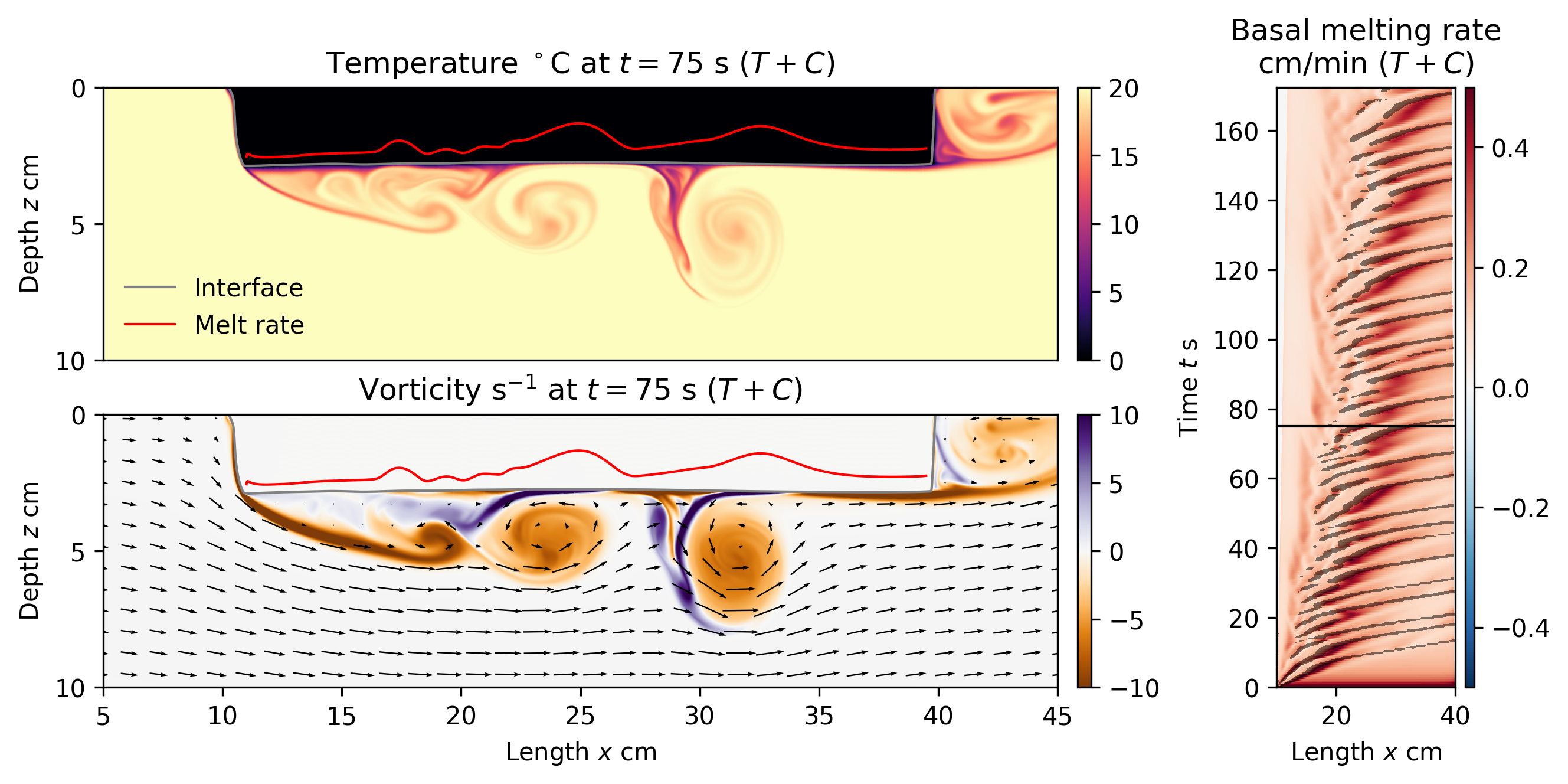}
	\caption{The left panels give snapshots of the temperature (top) and vorticity (bottom) for the $T+C$ simulation at \SI{75}{s}.
	The vorticity plot also shows the fluid velocity (black arrows).
	The qualitative melt rate (solid red line) is plotted relative to the ice-water interface, and is enhanced where vortices generate upwelling, with a maximum of \SI{0.35}{cm.min^{-1}} at $x=\SI{25}{cm}$ when $t = \SI{75}{s}$.
	The right panel plots the basal melt rate over space and time.
	Vortex locations (grey) are shown where the depth integrated vorticity is less than $-2$.
	The time $t=\SI{75}{s}$ is shown in black.}
	\label{fig:melting-flow-vortex-correlation}
	\end{figure}	

The first series simulates the experiments at high relative ambient velocity $U = \SI{3.5}{cm.s^{-1}}$.
The $T+C$ simulation uses the equation of state of seawater (EOS-80 \cite{FofonoffAlgorithmsComputationFundamental1983}) to capture salinity and buoyancy effects from experiments, while the $T$ only simulation ignores these effects. 

\Cref{fig:time-series-flow} shows a time series of temperature in the $T+C$ and $T$ only simulations.
Both simulations display similar patterns and reproduce several qualitative features of the experiment.
The largest melting occurs on the front (left) face, where warm ambient water collides with the ice.
The front melting increases with depth, causing a decrease in slope over time.
The base melts most rapidly at the centre, but is overall slower than the front.
These features are present in both simulations and therefore cannot be controlled by the buoyancy of the meltwater.
At high ambient velocities heat transported by the flow determines melting.

\Cref{fig:melting-flow} quantifies the front, basal, and rear melting for each simulation.
The front melting is steady in both simulations (\cref{fig:melting-flow} top left), with little deviation between instantaneous and cumulative melt rates (\cref{fig:melting-flow} bottom left), and an increase in front melting with depth (\cref{fig:melting-flow} bottom left).
The $T+C$ simulation reproduces the sloped front of experiments, while the $T$ only simulation has a vertical gradient at the top of the front.
This difference in slope occurs because cool, buoyant meltwater pools at the top of the $T+C$ simulation, thickening the thermal boundary layer and slowing melting.
This reduced melting also likely explains the small second peak in the histogram of front melt rates for the $T+C$ simulation in the second row of \cref{fig:melting-flow}.
Despite this difference, both simulations reproduce the experimental average front melt rate of $0.39\pm\SI{0.06}{cm.min^{-1}}$, with $0.46\pm\SI{0.14}{cm.min^{-1}}$ for the $T+C$ simulation, and 
$0.45\pm\SI{0.09}{cm.min^{-1}}$ for $T$ only (\cref{fig:melting-flow} centre left).

The basal melting shows significant variation in space and time (\cref{fig:melting-flow} top middle).
Both simulations show the same distinct regions of basal melting.
Just behind the leading edge, melt rates are low and steady.
This stagnant region is evident in the time series plots of \cref{fig:time-series-flow}.
Limited mixing with the warm ambient water causes reduced melting near the leading edge.
Behind this region, pooled meltwater becomes unstable due to the strong velocity shear, and by a mechanism akin to Kelvin-Helmholtz billowing, leads to vortex generation and shedding around the centre of the block.
The unsteady flow circulates warm ambient water to the base, and is associated with the transition to larger melt rates past the centre of the base.
Beyond this point the melting is on average lower, with intermittent periods of high melting occurring as vortices are shed downstream.
The transition to higher melting slowly moves backward over time.
This is partly because the front itself is receding, due to melting.
More importantly the slope of the front face reduces over time, delaying the instability of the shear layer generated at the separation point.
However the region of maximum cumulative melting does appear to saturate around $x = \SI{28}{cm}$ at late times.

The localised spatial and temporal features of the basal melting lead to nontrivial statistical properties, summarised in histograms of instantaneous melt rates in the second row of \cref{fig:melting-flow}.
There is large variance and skewness in the basal melt rates, as expected from the different melting regions.
The average basal melt rates of the simulations ($T+C: 0.17\pm \SI{0.08}{cm.min^{-1}}, T \text{ only}: 0.16 \pm \SI{0.08}{cm.min^{-1}}$) agree with the upper range of experimental results ($0.13 \pm \SI{0.04}{cm.min^{-1}}$), supporting the validity of the simulations.

The time averaged cumulative melt rates are given in row three, and clearly reproduce the localised increase in basal melting from the experiments.
The cumulative basal melt varies with distance by a factor of two, highlighting the importance of localised flow features in melting predictions.
Nevertheless it is clear that the buoyancy plays little role for large ambient velocities.
It is advection of heat that drives melting.

\Cref{fig:melting-flow-vortex-correlation} illustrates the process by which the flow induces melting hot spots.
The two left figures show snapshots of the temperature, vorticity, and velocity at $t = \SI{75}{s}$.
The temperature field shows a cool stagnant region that persists past the separation point at the front edge.
Mixing between this region and the ambient fluid is slow and intermittent, and the melt rate (red curve) remains low.
The vorticity plot shows prominent vortices being generated behind the stagnant region due to Kelvin-Helmholtz instability of the shear layer.
The fluid velocity (in arrows) reveals upwelling downstream from the vortices, which coincides with increased basal melting (red curve).
This explains both the mechanism by which vortices enhance melting -- upwelling of warm ambient water -- and accounts for the offset between vortex position and enhanced melting.

The right panel of \cref{fig:melting-flow-vortex-correlation} shows the instantaneous basal melt rate over space and time.
On top of this plot, we highlight the location of vortices in grey by thresholding the depth integrated vorticity.
This plot clarifies how vortices enhance melting.
Basal melt rates are low (roughly \SI{0.1}{cm.min^{-1}}) in the stagnant region where there are no vortices to generate mixing.
Behind this region vortices are being created.
These vortices transport warm ambient water to the base of the ice by upwelling to their right.
This is clear from the large melt rates (almost \SI{0.5}{cm.min^{-1}}) to the right of the vortex paths (grey stripes) in \cref{fig:melting-flow-vortex-correlation}. 
The vortices then reach a maximum size and are rapidly advected away by the flow.
In this stage they are no longer close enough to the base to enhance melting and the melt rate returns to a lower average value of around \SI{0.2}{cm.min^{-1}}.
We therefore do not expect a recurrence of localised melting without a new source of instability to generate vortices and enhanced mixing.

We note that the location of maximum melting is further downstream than observed in experiments.
This may be due to suppressed instability in the shallower simulation domain (15 cm vs 33.5 cm).
But the difference is most likely due to the behaviour of two-dimensional turbulence.
Whereas experiments show a rapid transition to turbulent mixing within \SI{10}{cm} of the front face, the two dimensional simulations take longer to develop instabilities, which manifest in larger coherent vortices in the flow.
While this does change the location of maximum basal melting, it does not change its magnitude.
We therefore expect this mechanism--increased upwelling at the transition to turbulence--to explain observed hot spots in experiments, and to persist for geophysical icebergs. 

The rear melting also shows noticeable variation in space and time (\cref{fig:melting-flow} top right), with an important difference from experiments.
While the rear face in experiments sloped to the right (\cref{fig:profiles}), the rear face of the $T+C$ simulation is on average vertical, and the $T$ only rear face slopes toward the left (\cref{fig:melting-flow} bottom right).
The incorrect sloping of the rear face in the simulations is likely due to large coherent vortices that remain behind the block, circulating warm ambient water toward the rear face from above (\cref{fig:time-series-flow}).
It is possible these coherent vortices are two-dimensional features which would break apart in three dimensions.
We also note very slight refreezing at the bottom, as cool meltwater is advected backward (\cref{fig:melting-flow} top right).
However, average rear melt rates are similar to the experimental value of $0.16 \pm \SI{0.07}{cm.min^{-1}}$, with $0.18 \pm \SI{0.13}{cm.min^{-1}}$ for the $T+C$ simulation, and $0.20 \pm \SI{0.11}{cm.min^{-1}}$ for the $T$ only simulation.
	
\subsection{No relative ambient velocity}
	\begin{figure}
	\centering
	\includegraphics[width=\linewidth]{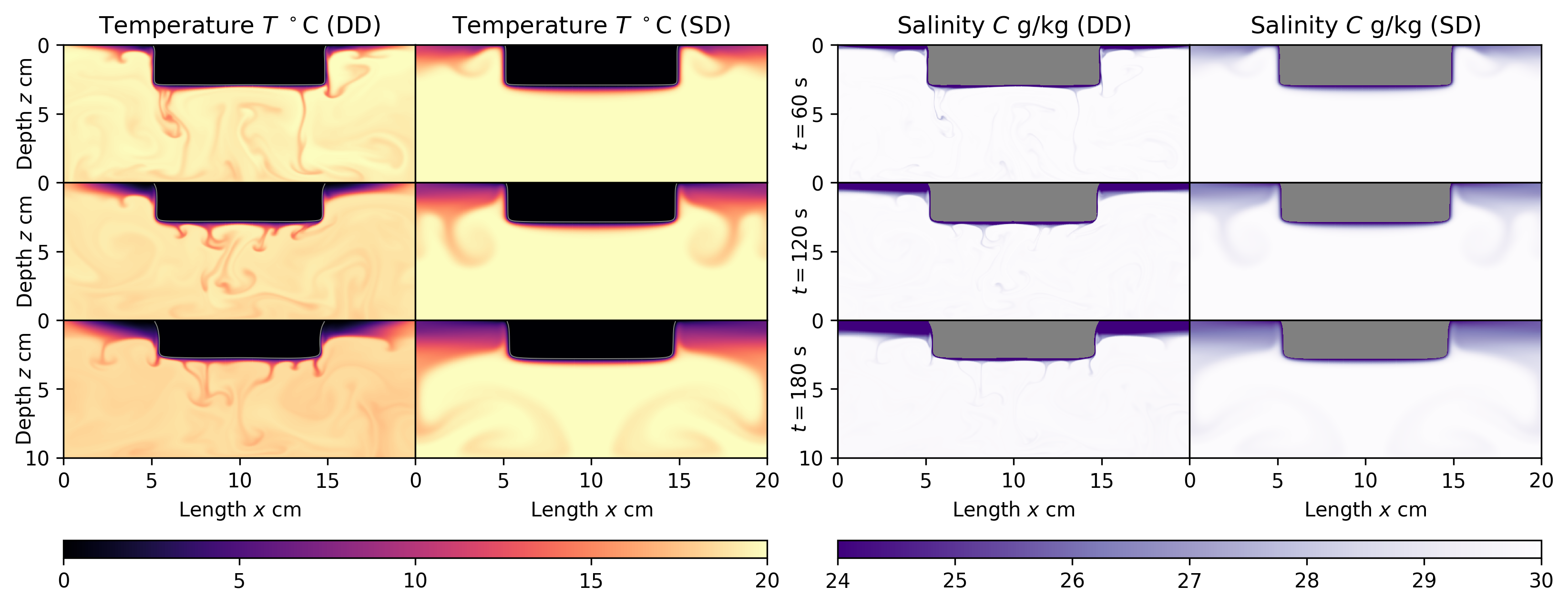}
	\caption{Time series, at one minute intervals, of temperature (left) and salinity (right) for simulations in quiescent salt water.
	The left simulation in each column uses a lower salt diffusivity (DD), while the right simulation in each column has equal salt and thermal diffusivity (SD).}
	\label{fig:time-series-noflow}
	\end{figure}
	
	\begin{figure}
	\includegraphics[width=.65\linewidth]{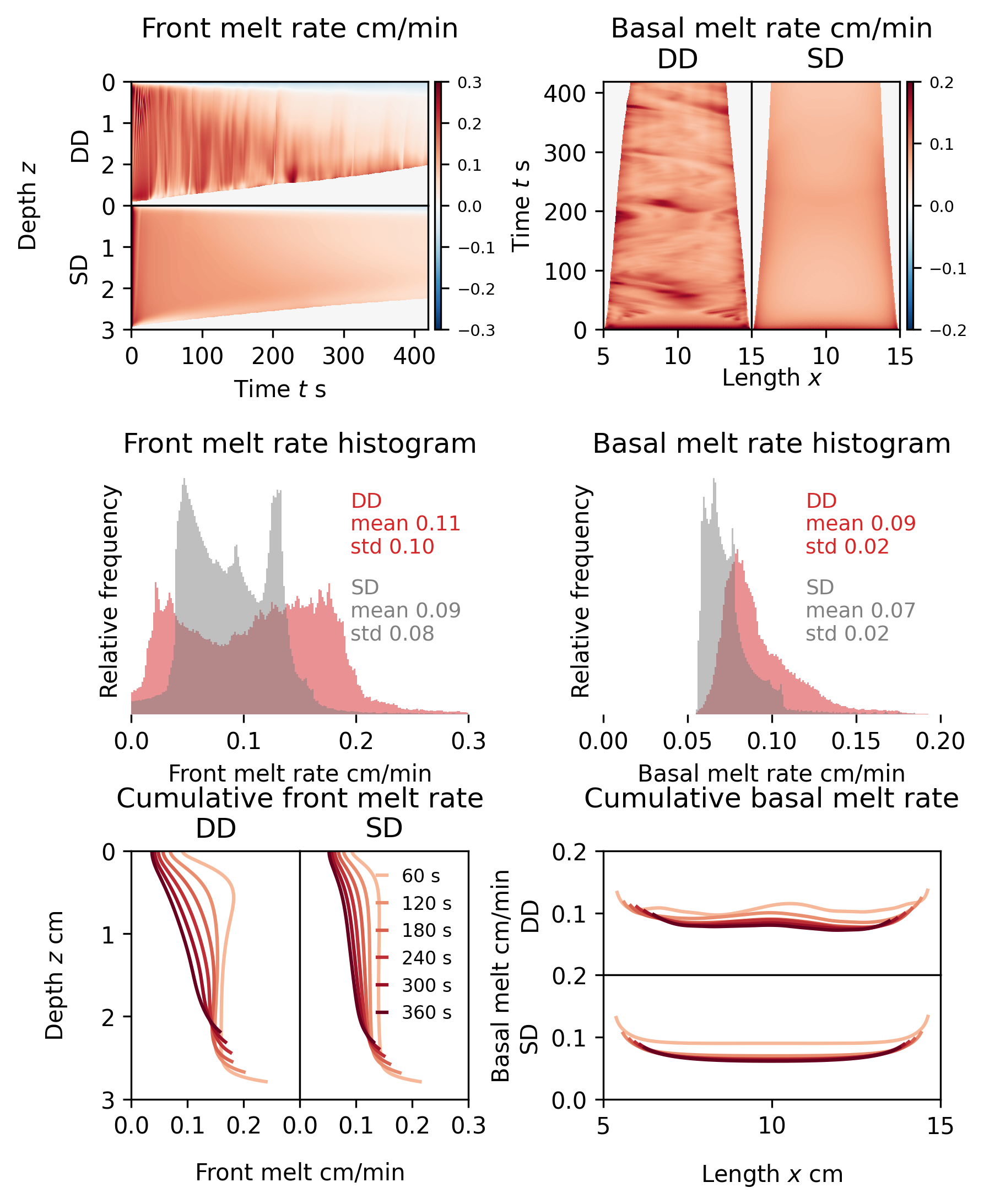}
	\caption{The left and right columns summarise side and basal melting respectively for the no relative ambient velocity simulations.
	The first row shows instantaneous melt rates over space and time.
	The second row gives a histogram, average, and standard deviation of instantaneous melt rates.
	The final row plots cumulative melt rate during the simulation.
	The right face is analogous to the left face and omitted.}
	\label{fig:melting-noflow}
	\end{figure}
	
The next simulation series investigates melting in an initially quiescent fluid $(U = \SI{0}{cm.s^{-1}})$.
Experiments showed some meltwater sinking beneath the block, despite the high salinity of the ambient water.
We show how double-diffusivity causes this sinking by comparing a double-diffusive (DD) simulation with different thermal and salinity diffusivity (Lewis number $\operatorname{Le}=\kappa/\mu=50/7$) to a single-diffusive (SD) simulation with equal diffusivities ($\operatorname{Le}=1$).
True Lewis numbers Le are much larger for salt water, however these simulations are enough to demonstrate the existence of double-diffusive effects. 
Both simulations use identical buoyancy functions (EOS-80 of seawater) with equivalent dependence on temperature and salt.

\Cref{fig:time-series-noflow} plots time series of temperature and salinity of the two simulations of melting ice in initially stationary salt water.
Both simulations show a clear tendency of meltwater (dark red/purple) to rise and pool near the free surface, reducing melting in that region at late times (\cref{fig:melting-noflow} top left).
But there are important differences between the simulations.
The layer of meltwater is more diffuse for equal diffusivities (SD).
And it is only with different diffusivities (DD) that sinking plumes emerge beneath the ice block.
This confirms the double-diffusive origin of sinking plumes in experiments.
Initially the sinking plumes are concentrated at the sides of the ice block, where the geometry aids the instability.
However the sinking tends over time to a persistent downward plume beneath the centre of the ice block (as in experiments).
In contrast, no sinking plumes occur beneath the ice block for the equal diffusivity simulation.
Some sinking is observed beneath the pooled meltwater, but this is because the rising water beneath the block sets up a recirculating flow that sinks and entrains meltwater near (periodic) horizontal boundaries.
All meltwater around the ice block rises without double-diffusion.
This leads to large differences in flow patterns, affecting melt rates.

\Cref{fig:melting-noflow} quantifies the melting of the two simulations.
The first row plots a colour time series of the instantaneous melt rates on the side and base of each block.
Fine-scale localised melting behaviour occurs on all faces in the double-diffusive simulation, whereas the equal diffusivity simulation shows little spatial or time variation.
This difference follows from the intermittent plumes generated via double diffusion.
The second row of \cref{fig:melting-noflow} shows that these distributions result from processes with nontrivial spatial and temporal structure, and are not simple Gaussian processes.
The mean and standard deviation of melt rates are larger for both faces for the double-diffusive simulation (DD).
The double-diffusive simulation has a side melt rate ($0.11\pm \SI{0.10}{cm.min^{-1}}$ (DD)) that is closer to experimental melt rates ($0.13\pm\SI{0.04}{cm.min^{-1}}$) than the equal diffusive simulation ($0.09\pm\SI{0.08}{cm.min^{-1}}$ (SD)).
The slight underestimate of side melting is an understandable consequence of the restricted domain size, which causes pooling meltwater to reduce side melt rates at late times.
This pooled melt water leads to some small amount of refreezing near the surface, visible in the top left panel of \cref{fig:melting-noflow}. 
The basal melting of the double-diffusive simulation ($0.09\pm \SI{0.10}{cm.min^{-1}}$ (DD))
and equal diffusivity simulation ($0.07\pm \SI{0.02}{cm.min^{-1}}$ (SD)) are both close to experimental basal melt rates ($0.08\pm \SI{0.016}{cm.min^{-1}}$).
In the final row (\cref{fig:melting-noflow}) we give the cumulative melt rates, which show faster side melting for the DD simulation at early times, which become slower at late times due to pooled meltwater.
Both simulations predict larger side melt rates than basal melt rates, as in experiments.
The difference in melt rates and flow patterns demonstrate the potential importance of double-diffusive effects for icebergs at low relative flow velocities.

\section{Geophysical application}
\label{sec:geophysical}
The laboratory and numerical results highlight that melt rates vary on each ice block face, and that different faces have different average melt rates.
The average melt rates of each face are also significantly higher than predicted by commonly used parameterisations.
These differences in melt rates matter for melting of real icebergs.

	\begin{figure}[ht]
    \centering
    \includegraphics[width=.8\linewidth]{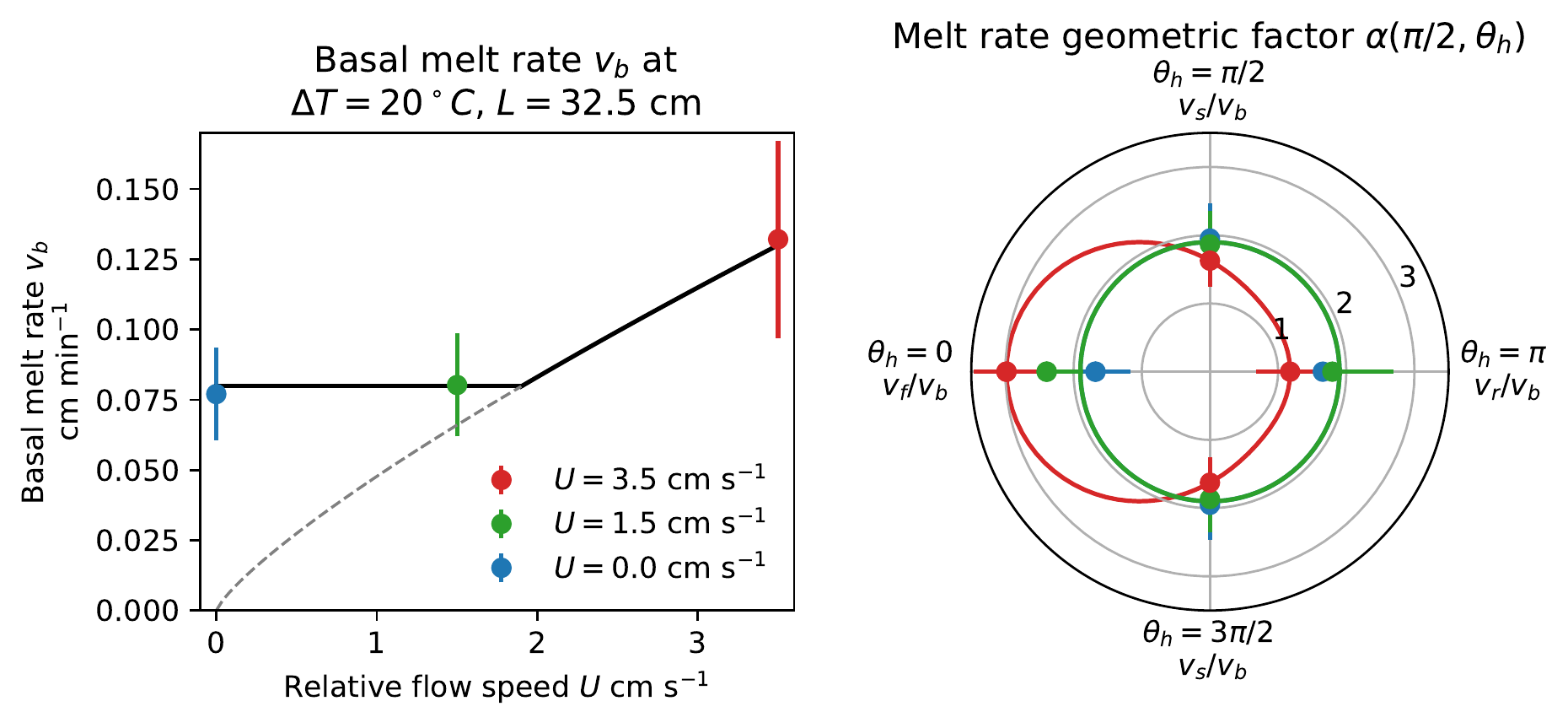}
    \caption{The left figure plots the experimental (coloured circles plus errors) and parameterised (black) basal melt rate $v_b$ as a function of ambient flow speed $U$.
    The grey dashed line shows the WC model below the transition speed $U^*=\SI{1.9}{cm.s^{-1}}$
    The right figure illustrates the experimental (coloured circles plus errors) and parameterised (solid) geometric factors $\alpha(\pi/2,\theta_h)$ for excess side melting at different horizontal orientations $\theta_h$ and flow speeds $U$.}
    \label{fig:parameterisation-figure}
	\end{figure}

We expect melt rates to differ between faces for real icebergs, as different faces will be exposed to different fluid velocities.
Even relatively low velocities towards an ice face cause higher melt rates than velocities parallel to the ice face \cite{JosbergerLaboratoryTheoreticalStudy1981}.
Different melt rates for different faces should occur even without relative ambient flows, as buoyancy driven plumes induce larger velocities and melt rates on the sides \cite{FitzMauriceNonlinearResponseIceberg2017}.
Furthermore, while absolute melt rates would be lower for the cooler waters of polar oceans (as melt rates are proportional to temperature), we expect geophysical icebergs to undergo the same overall melting patterns as the experiments.

To model different melt rates for each iceberg face, we propose an empirical parameterisation of iceberg melting that accounts for the orientation of the ice face in addition to the ambient flow speed $U$ and difference $\Delta T$ between the fluid and melting temperature.
The parameterisation decomposes the melt rate into a speed and temperature dependent parameterisation of basal melting $v_b(U,\Delta T)$ (using an adapted WC model similar to \cite{FitzMauriceNonlinearResponseIceberg2017}), and a multiplicative geometric factor $\alpha(\theta_v,\theta_h)$ that accounts for the orientation of each face,
	\begin{equation}
	v(U,\Delta T,\theta_v,\theta_h) = 
	v_b(U,\Delta T) \alpha (\theta_v, \theta_h).
	\end{equation}
Similar to \cite{FitzMauriceNonlinearResponseIceberg2017}, the basal melt rate $v_b$ at high flow speeds $U$ is modelled with a standard WC parameterisation.
The melt rate is rescaled by the factor $1.43$ to intercept the experimental basal melt rate at $U=\SI{3.5}{cm.s^{-1}}$. 
At low speeds, the basal melt rate $v_b$ is independent of the fluid velocity, and depends only on ambient temperature $\Delta T$,
	\begin{equation}
	v_b(U,\Delta T) = 
	\begin{cases}
	 	\SI{0.004}{cm.min^{-1}.\celsius^{-1}} \Delta T & U < U^*,\\
		1.43\cdot 0.037 \left(\frac{\rho_w}{\rho_i} \nu^{-7/15}\kappa^{2/3} \frac{c_p}{\Lambda}\right) \frac{U^{0.8}\Delta T}{L^{0.2}} & U \geq U^*.\\
	\end{cases}		
	\end{equation}
The transition between the regimes occurs for the speed $U^*$ at which the melt rates coincide.
For blocks with length $L = \SI{32.5}{cm}$, the transition occurs at $U^* = \SI{1.9}{cm.s^{-1}}$ (\cref{fig:parameterisation-figure} left).

The different melt rates of each face are accounted for in the geometric factor $\alpha(\theta_v,\theta_h)$.
Here, $\theta_v$ represents vertical inclination of the ice face (the angle between the inward-pointing ice face normal and the vertical unit vector), and $\theta_h$ represents the angle of the face relative to the ambient flow (the angle between the horizontal component of the inward-pointing ice face normal and the ambient flow velocity vector).
At all speeds, basal melting ($\theta_v = 0$) is less than melting on all vertical faces ($\theta_v = \pi/2$).
At low speeds ($U<U^*$), melt rates of all vertical faces are approximately equal, while at higher speeds ($U>U^*$), the side melting becomes anisotropic, with front melting ($\theta_h = 0$) exceeding side melting ($\theta_h = \pi/2$), which in turn exceeds rear melting ($\theta_h = \pi$).
The anisotropy increases linearly with the flow speed past $U^*$.
The geometric factor smoothly interpolates between experimental melt rates for each face using trigonometric polynomials in $\theta_v$ and $\theta_h$,
	\begin{equation}
	\alpha(\theta_v,\theta_h) = 
	\begin{cases}
 	1 + 0.9 \sin\theta_v & U < U^*,\\
 	1 + \sin\theta_v\br{0.9 + \frac{(U-U^*)}{\SI{3.5}{cm.s^{-1}} - U^*} \br{0.28 + 1.81 \cdot \frac{1}{2}\br{\cos\theta_h +1} - 0.45 \, (1 - \cos^2 \theta_h)}} & U \geq U^*.
	\end{cases}	
	\end{equation}
The basal melt rate $v_b$ for $\Delta T = \SI{20}{\celsius}$, $L = \SI{32.5}{cm}$ is plotted in the left panel of \cref{fig:parameterisation-figure}, and the geometric factor for each vertical face $\alpha(\pi/2,\theta_h)$ is illustrated in the right panel of \cref{fig:parameterisation-figure}.
This new parameterisation accounts for differences in average melt rates between iceberg faces.

Variations in melt rates matter because aspect ratios of geophysical icebergs also vary significantly, affecting the relative importance of melt rates on each face.
The B-15 iceberg had an estimated length of \SI{300}{km} and width of \SI{40}{km} \cite{ArrigoAnnualChangesSeaice2004}.
Using a depth of \SI{600}{m} as an upper limit \cite{DowdeswellKeelDepthsModern2007} suggests a minimum aspect ratio of approximately 180 (using the geometric mean of the two horizontal length scales).
At the other extreme, the Weeks stability criterion suggests a minimum aspect ratio of 1.4 for a \SI{50}{m} deep iceberg.
This means the basal area of B-15 comprises 97\% of the total submerged area while the basal area of a smaller marginally stable iceberg is only 27\%.

	\begin{figure}[ht]
    \centering
    \includegraphics[width=.5\linewidth]{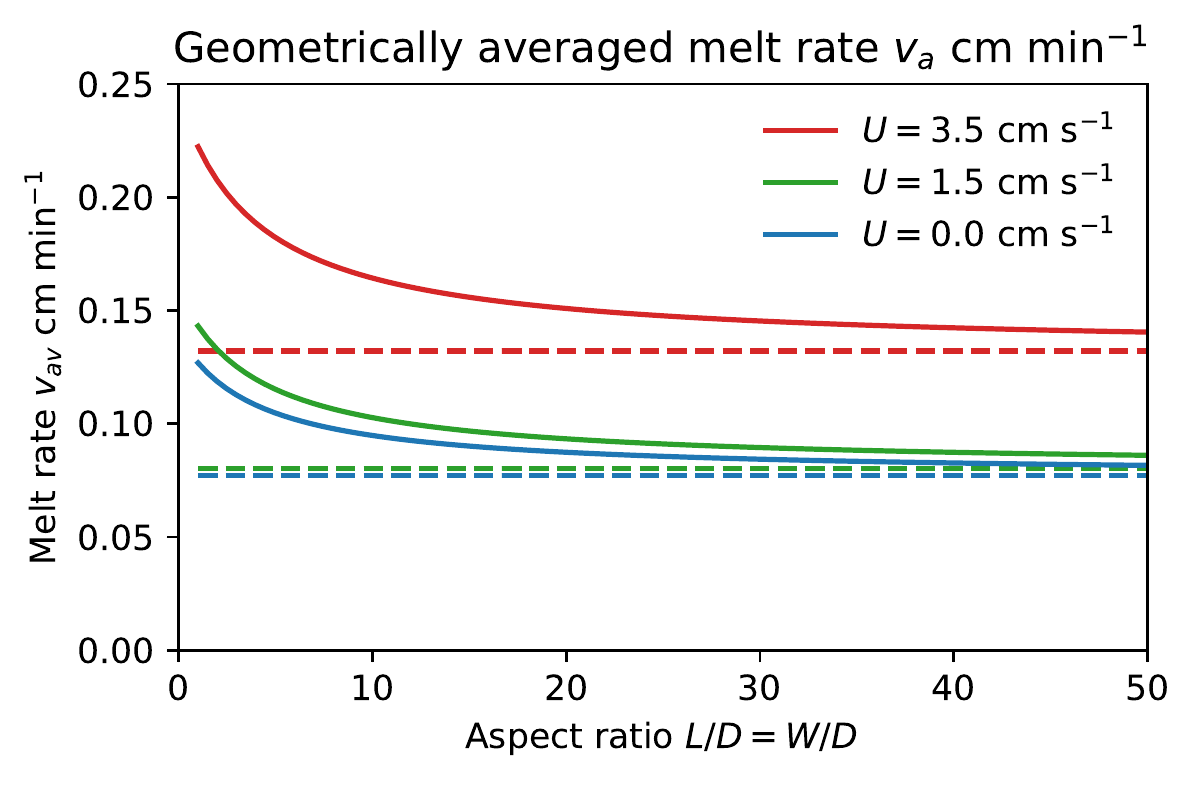}
    \caption{Illustration of geometry-weighted melt rates $v_{av} $ as a function of aspect ratio, for each relative ambient velocity $U$, using average experimental melt rates for each face (\cref{tab:melts}).
    The basal melt rate $v_b$ for each velocity is plotted in dashed lines.}
    \label{fig:aspect-ratio-figure}
	\end{figure}

We illustrate the influence of aspect ratio on melting using a simple geometric model of a melting block of ice that assumes each face melts uniformly.
The length $L$, width $W$, and submerged depth $D$ are related to melt rates of the front $v_f$, rear $v_r$, side $v_s$, and base $v_b$ via $\dot{L} = -v_f - v_r$, $\dot{W} = -2 v_s$, and $\dot{D} = -v_b$ (denoting time derivatives with dots).
The immersed volume $V$ and area $A$ are $V = LWD$, and $A = 2LD + 2WD + LW$.
The \emph{geometrically averaged} melt rate $v_{av}$ normalises the volume loss rate by the surface area,
	\begin{equation}
    v_{av} \equiv \frac{1}{A} \frac{d}{dt}|{V}| = (v_f+v_r)\frac{WD}{A}  + 2 v_s \frac{LD}{A} + v_b \frac{LW}{A},
	\end{equation}
which weights the melt rate of each face by its relative proportion of the total area.

\Cref{fig:aspect-ratio-figure} plots the geometrically averaged melt rate $v_{av}$ as a function of aspect ratio $L/D$, using average melt rates from experiments (\cref{tab:melts}) and simplifying the horizontal dimensions as equal ($W = L$).
\Cref{fig:aspect-ratio-figure} shows significant variation in the geometrically averaged melt rate $v_{av}$ from aspect ratio 1 to 50.
Ice blocks with unit aspect ratio $L=D$ melt more than 50\% faster than large aspect ratio ice blocks, with elevated overall melting (relative to the dashed basal melt rates) apparent even for aspect ratio 10.
We also note that the geometrically averaged melt rate is an instantaneous measure.
During melting the iceberg shape and hence geometrically averaged melt rate will evolve.

It is nevertheless straightforward to categorise the long term melting behaviour of icebergs as either side-dominated or base-dominated.
An iceberg will experience side-dominated melting if the aspect ratio decreases over time, and base-dominated melting if the aspect ratio increases over time.
If we simplify to two dimensions, with length $L$ and depth $D$, with respective melt rates $\dot{L} = 2v_s$ and $\dot{D} = v_b$, then the time derivative of the aspect ratio is
	\begin{equation}
	\frac{d}{dt}{\left(\frac{L}{D}\right)} = \frac{\dot{L} D - L \dot{D}}{D^2} = \frac{v_b}{D} \left(\frac{2v_s}{v_b} - \frac{L}{D}\right)
	\end{equation}
The time derivative of aspect ratio changes sign when the aspect ratio $L/D$ is equal to twice the melt rate ratio $2v_s/v_b$.
Therefore even icebergs with larger side melting can become dominated by basal melting if the aspect ratio is sufficiently large.
Iceberg B-15 would certainly experience base-dominated melting for realistic melt rate ratios.
And assuming that side melting is approximately twice as large as the basal melting, icebergs would transition from side-dominated to base dominated melting at aspect ratio 4.
Returning to three dimensions, if the melt rate differs between different side faces (as for the experiments with relative ambient velocity), different regimes can be defined for each orientation of the iceberg.

We also expect non-uniform melt rates on each face to affect geophysical icebergs.
The most common iceberg length of tabular icebergs in the Southern Ocean is \SI{400}{m}, and the most common freeboard $f$ is \SI{35}{m} \cite{RomanovShapeSizeAntarctic2012}. 
Using the empirical relationship $D = 49.4f^{0.2}$ \cite{RomanovShapeSizeAntarctic2012} gives a most common depth $D$ of approximately \SI{100}{m}.
Our experiments find a maximum in melting at a distance of two and a half times the depth, similar to studies of heat transfer past a forward facing step \cite{Abu-MulawehTurbulentMixedConvection2005}.
This predicts a maximum in basal melting at approximately \SI{250}{m}, a significant proportion of the basal length of the modal tabular iceberg.
This acts to further enhance overall melt rates for smaller aspect ratio icebergs.
Of course, confirming that this proportionality holds at larger scales would require further work.
Nevertheless studies of flow past a forward facing step find this length scale is a small multiple of the step height over a large range of Reynolds numbers \cite{SherryFlowSeparationCharacterisation2009}.

Non-uniform melting would also be enhanced by non-uniform background velocities, observed in Greenland fjords \cite{FitzMauriceEffectShearedFlow2016}, and induced by the Coriolis effects at large scales \cite{MeroniNonlinearInfluenceEarth2019}.
Ocean stratification would further alter velocity profiles of buoyancy induced flows.
We also note most iceberg shapes are irregular (up to 70\% non-tabular \cite{HotzelIcebergsTheirPhysical1983}) and that larger surface areas of irregular icebergs would further accelerate melting.
Improved iceberg melting parameterisations must therefore account for non-uniform melt rates induced by a range of physical effects.

\section{Conclusions}
\label{sec:conclusion}
Typical existing parameterisations of iceberg melting ignore the aspect ratio of icebergs \cite{WeeksIcebergsFreshWaterSource1973,HollandModelingThermodynamicIce1999}.
We conducted a series of laboratory experiments and numerical simulations to examine the dependence of the melt rate on iceberg size and shape for three different ambient velocities.
We find that geometry has a strong effect on iceberg melt rates.

Melt rates are highest on the forward facing side (with respect to the ambient flow), followed by the remaining lateral sides, with slowest melting occurring at the base of the iceberg.
We propose an empirical melt rate parameterisation that accounts for the orientation of each face, in addition to the ambient flow speed and temperature.
Using a toy model of iceberg melting, we show that changing the relative area of each face will thus change the overall melt rate,
with significant variation between small and large aspect ratio icebergs.

Furthermore, the melt rate of each face is itself spatially non uniform, with localised increases in basal melt rates of up to 50\% observed.
These localised regions correspond to the reattachment zones of flow past a forward-facing block, and occur at a distance approximately two to three times the depth of the block \cite{Abu-MulawehTurbulentMixedConvection2005}.
Our numerical investigation showed that these non-uniformities in basal melt are caused by the generation of vortices, which lead to upwelling of warm water during their formation.
This non-uniformity exists at large relative speeds, and is not influenced by buoyancy.
We therefore expect these non-uniformities to increase melting for small-aspect ratio icebergs.

To improve melting estimates, we emphasise that models of melt rates must depend on both the speed and orientation of the background flow, and that differing melt rates must be weighted according to the shape and aspect ratio of an iceberg.

\subsection{Acknowledgments}
Eric Hester is grateful for support from NSF OCE-1829864 during his 2017 Geophysical Fluid Dynamics Summer Fellowship at the Woods Hole Oceanographic Institution, as well as support from the University of Sydney through the William and Catherine McIllarth Research Travel Scholarship.
Louis-Alexandre Couston acknowledges funding from the European Union's Horizon 2020 research and innovation programme under the Marie Sklodowska-Curie grant agreement 793450. 
Claudia Cenedese was supported by NSF OCE-1658079.
We acknowledge PRACE for awarding us access to Marconi at CINECA, Italy.


\newcommand{\noopsort}[1]{}

\end{document}